\documentclass{aa}
\usepackage{graphicx}
\usepackage{txfonts}
\usepackage{epsfig}
\def\ltsima{$\; \buildrel < \over \sim \;$}
\def\gtsima{$\; \buildrel > \over \sim \;$}
\def\simlt{\lower.5ex\hbox{\ltsima}}
\def\simgt{\lower.5ex\hbox{\gtsima}}

\begin{document}

   \title{The $Suzaku$ X-ray spectrum of NGC~3147. 
Further insights on the best ``true'' Seyfert 2 galaxy candidate}

   \author{G. Matt \inst{1},  S. Bianchi\inst{1},  M. Guainazzi\inst{2}, 
X. Barcons\inst{3}, F. Panessa\inst{4}
}

   \offprints{G. Matt, \email{matt@fis.uniroma3.it} }

   \institute {$^1$Dipartimento di Fisica ``Edoardo Amaldi'', Universit\`a degli Studi Roma Tre, 
via della Vasca Navale 84, I-00146 Roma, Italy \\
$^2$ European Space Astronomy Center of ESA, Apartado 50727, 28080 Madrid, Spain \\
$^3$ Instituto de Fisica de Cantabria (CSIC-UC), 
39005 Santander, Spain\\
$^4$  Istituto di Astrofisica Spaziale e Fisica Cosmica (IASF-INAF), 
via del Fosso del Cavaliere 100, 00133 Roma, Italy
}

   \date{Received / Accepted }

   \abstract
{NGC 3147 is so far the most convincing case of a ``true'' Seyfert 2 galaxy, i.e.
a source genuinely lacking the Broad Line Regions.}{We obtained a $Suzaku$
observation with the double aim to study in more detail the iron line complex, and to check
the Compton-thick hypothesis for the lack of observed
optical broad lines.}
{The $Suzaku$ XIS and HXD/PIN spectra of the source were analysed in detail.}{The line
complex is composed of at least two unresolved lines, one at about 6.45 keV 
and the other one at about 7 keV, most likely identified with Fe XVII/XIX, the former, 
and Fe XXVI, the latter. The high-ionization line
can originate either in a photoionized matter 
or in an optically thin thermal plasma. In the latter case, 
an unusually high temperature is implied. In the photoionized model case, 
the large equivalent width can be explained 
either by an extreme iron overabundance or by assuming that the source is Compton-thick. In the
Compton-thick hypothesis, 
however, the emission above 2 keV is mostly due to a highly ionized reflector, contrary to
what is usually found in Compton-thick Seyfert 2s, where reflection from low ionized matter 
dominates. Moreover, the source flux varied between the XMM-$Newton$ and the $Suzaku$ 
observations, taken 3.5 years apart, confirming previous findings and 
indicating that the size of the emitting region must be
smaller than a parsec.
 The hard X-ray spectrum is also inconclusive on the Compton-thick hypothesis. Weighting
the various arguments, a ``true'' Seyfert 2 nature of NGC~3147 seems to be still
the most likely explanation, even if the ``highly 
ionized reflector'' Compton-thick hypothesis cannot at present be formally rejected.}{}

\authorrunning{G. Matt et al. }
\titlerunning{The $Suzaku$ X-ray spectrum of NGC~3147}

   \maketitle
%

\section{Introduction}

NGC~3147 ($z$=0.009346) is at present the most convincing example
of a ``true'' Seyfert 2 galaxy (Bianchi et al. 2008; Shi et al. 2010;
Tran et al. 2011). In such sources
the lack of detection of broad emission lines in the optical/UV
spectra cannot be explained by the obscuration of the Broad Line Regions (BLR) - as it 
is the case for ``normal'' Seyfert 2s in the so far very successful 
Unification models (Antonucci 1993) - but requires their absence. 

In fact, {\it simultaneous} X-ray and optical
spectroscopy have demonstrated that in this source
the BLR is lacking at the same
time when the nucleus is seen unobscured (Bianchi et al. 2008). 
The X-ray XMM-$Newton$ observation, albeit rather short (less than 20 ks), 
was good enough to put a tight constraint
on the column density of any obscuring material along the line-of-sight. The
same observation also indicated complex iron line emission, but the short
exposure time did not permit a detailed analysis. 

It has been suggested (Nicastro 2000; Nicastro et al. 2003; Elitzur \& Shlosman 2006)
that at low accretion rates (and thence low luminosities)
the broad line regions (BLR) cannot form.
Indeed, NGC~3147 is a low (even if not extremely so) luminosity source 
accreting at a low rate (bolometric luminosity 
$L\sim 5\times10^{42}$ erg s$^{-1}$, $L/L_{Edd}\sim 10^{-4}$, Bianchi et al. 2008).

A possible alternative explanation for the peculiar characteristics of NGC~3147
is that the source is Compton-thick. In fact, the simultaneous lack 
of BLR and absorption can be explained if the absorber is so thick to
completely obscure the nuclear emission below 10 keV. In this case,
the observed emission should be due to reflection off circumnuclear matter,
but X-rays could pierce through the absorbers above 10 keV.
The hard X-ray coverage offered by $Suzaku$ is the best tool at present to test it.

With the dual goal of studying in more detail the iron line emission and to
check the Compton-thick explanation for this 
source, we asked for, and obtained, a long (150 ks) 
$Suzaku$ observation. 

The paper is organized as follows. The observation and data reduction
are described in Sec.~2, while the data analysis is presented in Sec.~3. 
Results are discussed and summarized in Sec.~4.

\begin{figure}
\epsfig{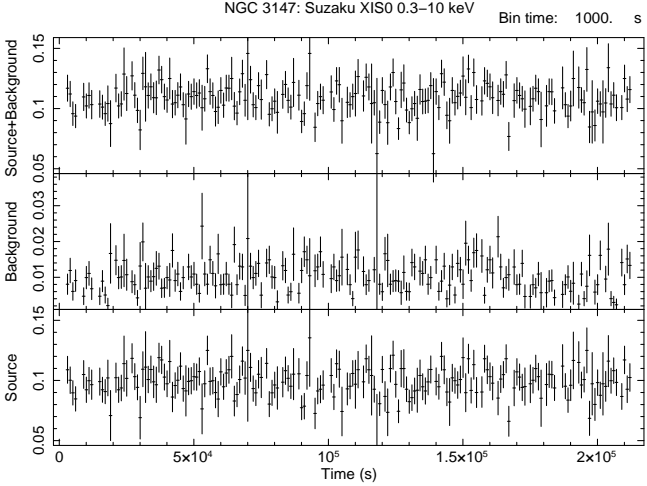}
  \caption{XIS0 light curve. Upper panel: count rates
in the source extraction region. Medium panel: count rates
in the background. Lower panel: background-subtracted source
count rates.
}
  \label{lcurve}
\end{figure}

\section{Observation and data reduction}

NGC~3147 was observed by \textit{Suzaku} on 2010, May 24, for 150 ks (\textsc{obsid} 705054010).
X-ray Imaging Spectrometer (XIS) and Hard X-ray Detector (HXD) event files were
reprocessed with the latest calibration files available (2011-06-30 release),
using \textsc{ftools} 6.11 and \textit{Suzaku} software Version 18, adopting
standard filtering procedures. Source and background spectra for all the three
XIS detectors were extracted from circular regions with radius of 167 pixels
($\simeq175$ arcsec), avoiding the calibration sources. Response matrices and
ancillary response files were generated using \textsc{xisrmfgen} and
\textsc{xissimarfgen}. We downloaded the ``tuned'' non-X-ray 
background (NXB) files for
our HXD/PIN data provided by the HXD team\footnote{see 
ftp://legacy.gsfc.nasa.gov/suzaku/data/background/pinnxb\-\_ver2.0\_tuned/}  
and extracted source and background
spectra using the same good time intervals. The PIN spectrum was then corrected
for dead time, and the exposure time of the background spectrum was increased by
a factor of 10, as required. Finally, the contribution from the cosmic X-ray
background (CXB) was subtracted from the source spectrum, simulating it as
suggested by the HXD team.

The XIS spectra
were fitted between 0.5 and 10 keV.
The normalizations of XIS1 and XIS3 with respect to XIS0 were left free in the fits,
and always resulted in agreement to within 3\%. A constant factor of 1.16 was
used instead between the PIN and the XIS0, as recommended for observations taken
in the XIS nominal position. In the following, all the PIN fluxes are given
with respect to the XIS0 flux scale, which is 1.16 times 
lower than the HXD absolute flux scale.

In the following, quoted statistical errors correspond to the 90\% confidence level for one
interesting parameter ($\Delta\chi^2=2.71$), unless otherwise stated. The
adopted cosmological parameters are H$_{0}=70$ km s$^{-1}$ Mpc$^{-1}$,
$\Omega_\Lambda=0.73$, and $\Omega_m=0.27$ (i.e., the default ones in
xspec 12.7.0: Arnaud 1996). We use the Anders \& Grevesse (1989) 
chemical abundances and the photoelectric absorption cross-sections by 
Balucinska-Church \& McCammon (1992). 

\section{Data analysis and results}

The light curve is shown in Fig~\ref{lcurve}. No significant
variability is detected. For the spectral analysis, therefore, 
we used spectra integrated over the whole observation.

\subsection{The XIS spectra}

\begin{figure}
\epsfig{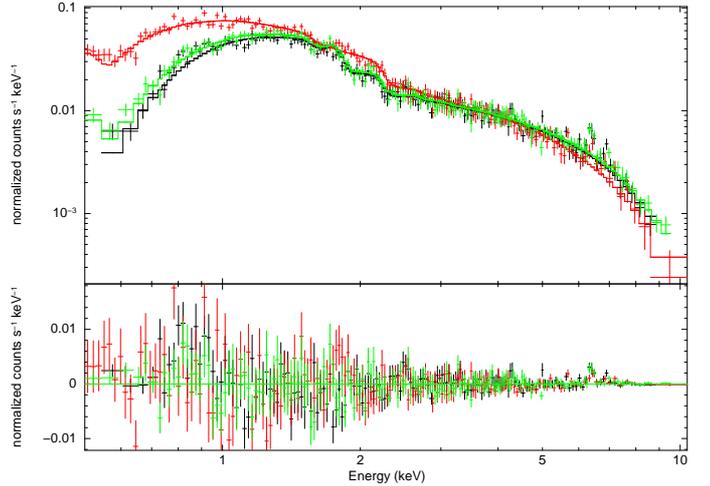}
  \caption{Best fit model and residuals fitting the XIS spectra
with a simple power law plus Galactic absorption.
}
  \label{bestfit}
\end{figure}

We first analyzed the XIS spectra. A simple power law model 
(plus Galactic absorption, N$_{H,G}$=3.64$\times10^{20}$ cm$^{-2}$, 
Dickey \& Lockman 1990) gives a poor fit ($\chi^2/d.o.f.$=604.1/482).
The most prominent features in the residuals are apparent
at the energies of the iron line complex (see Fig.~\ref{bestfit}). 
We therefore added two narrow emission lines, with
energies free to vary around 6.4 and 7 keV (see next section), which 
improves significantly the quality of the fit ($\chi^2/d.o.f.$=496.4/478). 
The power law index is 1.745$\pm0.014$.
The inclusion of either a warm absorber or a Compton reflection component 
is not required by the data ($\Delta\chi^2$=-2.4 and -2.8, respectively), 
while addition of a second power law improves the fit quality
($\chi^2/d.o.f.$=481.3/476; the improvement is significant at the 99.94\%
confidence level, according to the F-test). The power law indices
are 3.50($^{+0.42}_{-0.69}$) and 1.689($^{+0.022}_{-0.018}$). The 0.5-2 
(2-10) keV flux is 8.39$\times$10$^{-13}$ (1.64$\times$10$^{-12}$)
erg cm$^{-2}$ s$^{-1}$, corresponding to a (absorption corrected)
luminosity of 1.8$\times$10$^{41}$ (3.2$\times$10$^{41}$) erg s$^{-1}$. 

Finally, no intrinsic
absorption is detected. The upper limit to the column density of any
neutral absorber at the redshift of the source is 5$\times10^{20}$ cm$^{-2}$.

\subsubsection{Comparison with the XMM-$Newton$ results}

We then compared the $Suzaku$ spectrum with the XMM-$Newton$
one (Bianchi et al. 2008). XMM-$Newton$ observed NGC~3147 on
2006, October 6 (i.e. about three and half year earlier), with
a net exposure time of 14 ks in the EPIC-pn instrument.
We re-extracted the spectrum with the same procedure described
in Bianchi et al. (2008), but using the latest version of SAS
(11.0.1) and of the calibration files.

According to Bianchi et al. (2008), we fitted the spectrum with
a power law absorbed by both Galactic and intrinsic matter, and two
narrow guassian lines. The fit is good ($\chi^2/d.o.f.$=93.6/106).
The power law index is 1.62$\pm$0.05, the
column density of the local absorber is $<$3.2$\times10^{20}$ cm$^{-2}$, 
the energies of the two lines are 6.471($^{+0.089}_{-0.065}$) keV and
6.798($^{+0.083}_{-0.072}$) keV and their fluxes are 2.8($\pm$1.6)$\times10^{-6}$
ph  cm$^{-2}$ s$^{-1}$ (EW of 189 eV) and  2.4($\pm$1.5)$\times10^{-6}$ ph  cm$^{-2}$ s$^{-1}$
(EW of 172 eV), respectively. All these values are consistent within the errors
with those of Bianchi et al. (2008). The 0.5-2 
(2-10) keV flux is 5.78$\times$10$^{-13}$ (1.43$\times$10$^{-12}$)
erg cm$^{-2}$ s$^{-1}$.

\begin{figure}
\epsfig{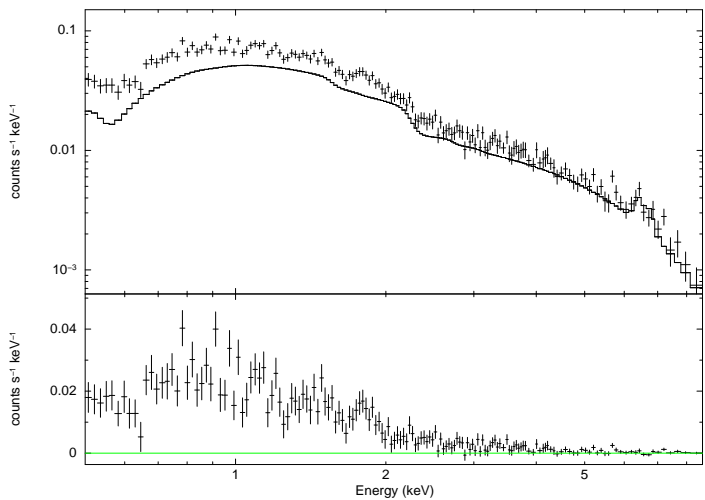}
\label{XIS_XMM}
 \caption{The $Suzaku$/XIS1 spectrum and residuals when the best fit
model for the XMM-$Newton$ observation is superimposed.
}
\end{figure}

Comparing the XMM-$Newton$ and $Suzaku$ results, a variation, most
prominent in the soft band, is found. This is shown in 
Fig~\ref{XIS_XMM}, where the XMM-$Newton$ best fit model is superposed
to the Suzaku/XIS1 spectrum. The variation can be explained either as
a steepening of the power law component or, better, with 
different variations of the soft and hard X-ray component, the former
varying most. An explanation purely in terms of a variation of the 
local absorber is, instead, not viable. 

This result confirms that the source is variable on
time scales of years. In fact, the measured 2-10 keV flux, in units
of 10$^{-12}$ erg cm$^{-2}$ s$^{-1}$, has been measured as 1.6 in September 
1993 (with ASCA: Ptak et al. 1996), 2.3 in November 1997 (with BeppoSAX: Dadina 2007)
and 3.7 in September 2001 (with Chandra: Terashima \& Wilson 2003). That the 
variation cannot be due to a confusing source is demostrated by the fact
that the highest flux has been measured by the best spatial resolution
satellite.

\subsection{The iron line complex}

\begin{figure}
\epsfig{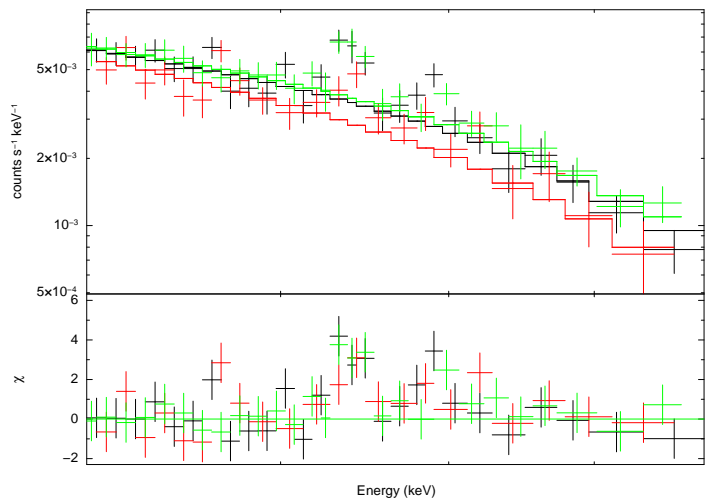}
  \caption{Data and best fit model between 5 and 9 keV. The model is the 
best fit one (see text), but without any emission lines. Line emission
around 6.4 keV and 7 keV is apparent. 
}
  \label{ironline}
\end{figure}

We then studied in more detail the iron line complex.  While the
iron line properties in the XMM-$Newton$ and $Suzaku$ 
observations are roughly consistent, 
the much better statistics in the $Suzaku$ observation
due to the longer exposure time allows for a more detailed analysis.
To this end, we limited for simplicity the analysis to the XIS instruments,
and the energy range to the 4-10 keV band. In Fig.~\ref{ironline},
the data and best fit model are shown, after removal of any emission line
from the fit. The presence of two emission lines is clear, one 
around 6.4 keV, to be attributed to K$\alpha$ emission from 
neutral or low ionization iron, the other
around 7 keV, likely due to either (or both) hydrogen-like (i.e. Fe XXVI) iron emission or
K$\beta$ emission from low ionization iron. We therefore fitted the spectrum 
with a model composed by a single power law plus three narrow ($\sigma$=0)
emission lines, with rest-frame energies fixed at 6.4 keV
 (neutral iron, K$\alpha$), 6.96 keV 
(hydrogen-like iron, K$\alpha$), and 7.06 keV (neutral iron, K$\beta$). The fit is good
($\chi^2/d.o.f.$=116.2/120), but some wiggles are apparent in the residuals
around the K$\alpha$ neutral iron line. Letting the energy of that line free to 
vary, a significantly better fit is found ($\chi^2/d.o.f.$=99.5/119), with a line
energy of about 6.45 keV. No improvement, instead, is found letting
the width of that line free to vary ($\sigma<$63 eV), nor including the
He-like iron line (upper limit to the flux of 5.5$\times10^{-7}$ ph cm$^{-2}$ s$^{-1}$, 
corresponding to an equivalent width of 35 eV). Not surprisingly, given the narrowness
of the line, no improvement is found with a relativistic profile ({\sc diskline} model),
and the inner disc radius is very large (hundreds of gravitational radii). Even in the
relativistic line model, an intrinsic line energy at 6.45 keV is strongly preferred by
the fit. 

\begin{figure}
\epsfig{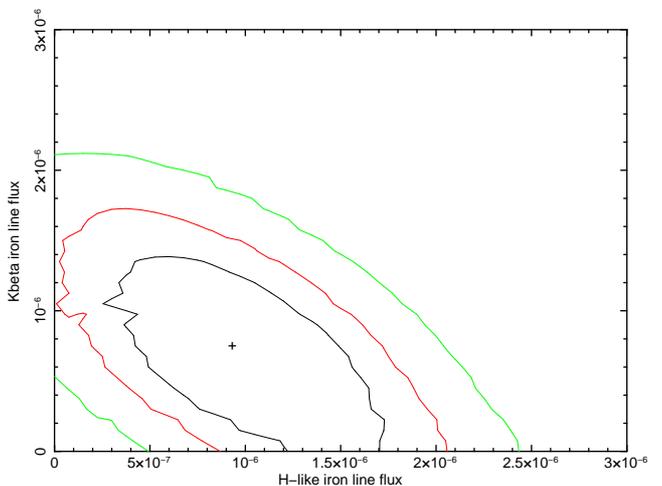}
  \caption{Contour plot of the K$\beta$ low-ionization
line flux vs. the hydrogen-like K$\alpha$ line flux. 
}
  \label{contour_iron}
\end{figure}

The results are summarized in Table~\ref{lines}. The line fluxes are consistent
within the errors to those derived from the XMM-$Newton$ spectra.
 While the hydrogen-like line is only marginally detected
and the K$\beta$ line is formally an upper limit, this of course 
does not mean that there is no significant line emission
at that energy, but simply that the quality of the data is not good enough
to accurately determine the parameters of the lines simultaneously.
This is best seen in Fig.~\ref{contour_iron}, where the contour
plot of the fluxes of the two lines is shown: a
simultaneous lack of emission for the two lines is in fact not allowed. 

\subsubsection{The low ionization iron line}

The rest-frame energy of the low ionization 
K$\alpha$ iron line is significantly larger than 
(and not consistent with) 6.4 keV, the energy for neutral iron. 
We are not aware of any major problem in the
energy calibration of the XIS detectors and indeed, fitting the Mn K$\alpha$ doublet and the K$\beta$
in the calibration spectra, we found energies consistent with the intrinsic ones. 
Letting the energy of the line free to
vary independently in the three XISs, similar values are found. We therefore conclude
that the iron is truly ionized. 

According to House (1967), the best-fit energy corresponds to
Fe XVII/XIX, where iron emission should be suppressed by resonant trapping (Ross \& Fabian
1993,  Matt et al. 1993, 1996). To avoid this effect, the matter should be very optically
thin, which is ruled out by the large EW (almost 200 eV), or very turbulent, so reducing
the effective optical depth at the line core. Recent and more refined calculations 
by Garcia et al. (2011), however, show that suppression is not so efficient and that
significant emission from the abovementioned ions is possible. In this case, we do not
expect any K$\beta$ line emission, because for those ions the M shell is no longer populated,
and therefore the line emission at about 7 keV should be entirely due
to the hydrogen-like iron line. 

\subsubsection{The high ionization iron line}

Setting the K$\beta$ line to zero, the H-like line flux is 
1.44($\pm$0.59)$\times$10$^{-6}$  ph cm$^{-2}$ s$^{-1}$ (see also Fig.~\ref{contour_iron}),
with an equivalent width of about 100 eV\footnote{Again, assuming a relativistic profile
a fit as good as the one with a gaussian line is obtained, but with a very
large inner radius.}, and $\chi^2/d.o.f.$=101.5/120.
 An upper limit of 35:100 to the ratio between helium- and hydrogen-like lines 
(as derived from the respective fluxes) 
implies high values of the ionization parameter, such that the EW of the hydrogen line should
be, for a solar iron abundance, only a few eV with respect to the total 
continuum (Bianchi \& Matt 2002). The required
factor of (at least) 10-20 iron overabundance is rather extreme, even if not fully 
inconsistent with the EW of the low ionization line (which is 
a factor 2-3 larger than expected  for solar abundances, 
in agreement with the calculations of Matt et al. 1997 for
optically thick matter with a similar iron overabundance). 

On the other hand, such large equivalent widths {\bf could} be explained if the source is Compton-thick
and the observed emission comes from reflection in two physically distinct
mirrors, one at low and the other at high ionization, and no direct emission.
To test this hypothesis, we fitted the spectrum with a power law (representing
reflection from highly ionized matter), a Compton reflection component (to account for reflection 
from low ionized matter), plus two gaussian lines, one free to vary around 6.4 keV and the other
fixed to 6.96 keV. The value of $R$, the relative amount of Compton reflection with respect
to the power law component, was fixed to 2, which is the value corresponding to the 
observed EW of the 6.4 keV line. This translates to a 2-10 keV flux from the high
ionization reflector 7 times larger than the low ionization component,
a quite unusual configuration in Compton-thick sources where it is the
low ionization reflection which generally dominates. The fit is good 
($\chi^2/d.o.f.$=107.8/120), even if
slightly worse than with a simple power law. The H-like line flux is now 
0.97($^{+0.77}_{-0.60}$)$\times$10$^{-6}$  ph cm$^{-2}$ s$^{-1}$, corresponding to an EW with respect
to the high ionized reflection only of about 80 eV, a value consistent with what is expected from the 
large ionization parameter values implied by the absence of the He-like line (Bianchi \& Matt 2002).

Alternatively, this line may be emitted in a high temperature, optically thin thermal plasma.
A 0.5-10 keV fit with the {\sc mekal} model (iron abundance fixed
to solar) instead of the 7 keV gaussian line is acceptable ($\chi^2/d.o.f.$=485.5/476), 
giving a plasma temperature of 25($^{+23}_{-5}$) keV. The 2-10 keV luminosity of
the thermal component is 1.2$\times$10$^{41}$ erg/s. The emission measure, $\sim \int n_H^2dV$,
is about 7$\times$10$^{63}$ cm$^{-3}$. Assuming, for self-consistency, that the emitting
matter is not optically thick to Thomson scattering, i.e. $n_HR\sigma_T<$1 (with $R$ the radius
of the emitting region, assumed spherical and with constant density), then a
lower limit to $R$ of 7.4$\times$10$^{14}$ cm (i.e. about 12 Schwarzschild radii
if a value of the black hole mass of 2$\times$10$^{8}$ solar masses is adopted, Dong
\& DeRobertis 2006), is obtained (corresponding to an upper 
limit to $n_H$ of 2$\times$10$^{9}$ cm$^{-3}$). 
(If the {\sc mekal} component is used instead of both the hard power law 
and the highly ionized iron line
the fit is still good, the soft power law gets flatter,  $\Gamma$=2.18($^{+0.13}_{-0.21}$)
and the thermal plasma 2-10 keV luminosity is 2.2$\times$10$^{41}$ erg s$^{-1}$).

The origin and nature of this putative thermal plasma is however unclear. 
The accretion rate of NGC~3147 is low ($L/L_{Edd}\sim 10^{-4}$, Bianchi et al. 2008), and
therefore accretion could occur in a radiatively inefficient mode
(ADAF, Narayan et al. 1995; RIAF, Yuan et al. 2003), where the X-ray emission is
expected to be due to bremsstrahlung radiation. However, in such a mode 
the ion temperature is expected to be extremely high, and no visible line
is expected. The thermal emission may be related to hot gas in a starburst region,
but the implied luminosities are quite large, and there is no evidence for such an
extreme starburst at other wavelengths. Alternatively, the plasma emitting region may be
a compact one, in the innermost regions of the AGN.  
While there is so far no strong evidence of a significant contribution of such a 
plasma emission to the X-ray spectrum of Seyfert galaxies, 
it must be recalled that NGC 3147 is likely a peculiar source.


\begin{table}
  \caption{Best-fit parameters for the iron lines.}
  \begin{tabular}{cccc}
    \hline
& & & \\     
    & E   & Flux & EW   \\
    & (keV) & (10$^{-6}$ ph cm$^{-2}$ s$^{-1}$) & (eV)  \\
& & & \\     
\hline
& & & \\     
    Low ionized $K\alpha$  & 6.452$^{+0.021}_{-0.015}$ & 3.27$\pm$0.61 & 195$\pm$36 \\ 
    H-like &  6.96  &  0.93$\pm$0.89 & 63$\pm$60 \\
    Low ionized K$\beta$ & 7.06  & $<1.61$ & $<$111 \\
& & & \\     
    \hline
  \end{tabular}{\noindent}
\label{lines}
\end{table}

\begin{figure*}
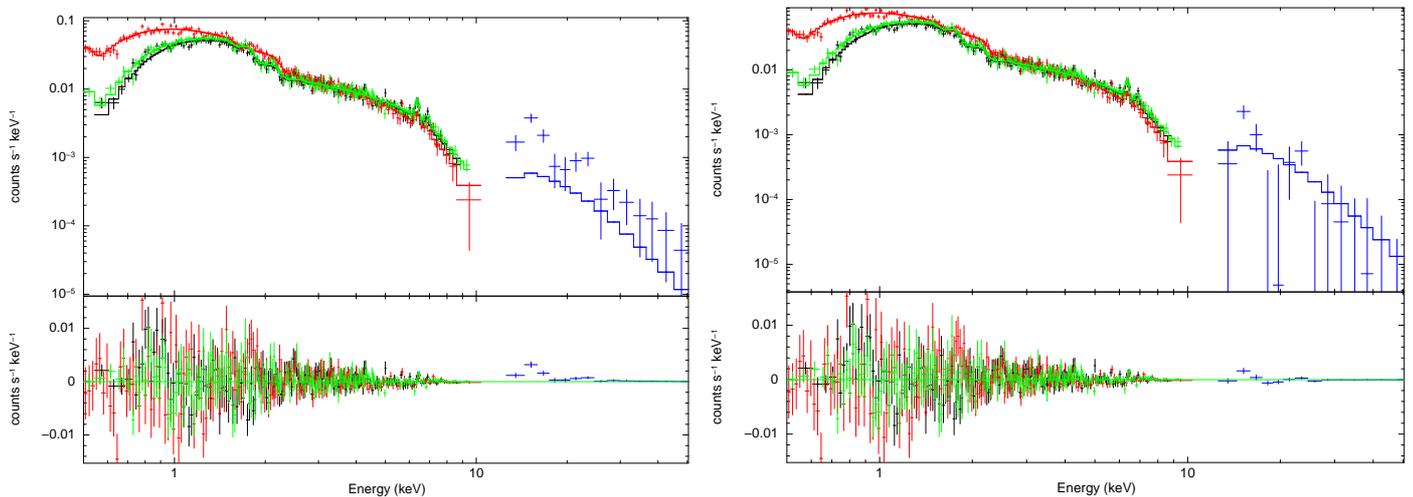

\epsfig{file=plot_pin.ps,width=6.5cm,angle=-90}
\hfill
\epsfig{file=plot_pincor0035.ps,width=6.5cm,angle=-90}
  \caption{XIS and HXD/PIN spectra fitted with the best
fit XIS-only model. A clear excess at hard X-ray
is observe (left panel), which however disappears (right panel)
once a 3.5\% systematic increase in the background is included
(see text for discussion).
}
  \label{pin}
\end{figure*}

\subsection{The hard X-ray emission}

One of the goals of the $Suzaku$ observation of NGC~3147 is to exploit
its hard X-ray coverage to 
test the Compton-thick hypothesis for the unabsorbed X-ray spectrum
of this Seyfert 2 galaxy. In fact, if the source is Compton-thick,
the spectrum may appear unabsorbed because the primary emission 
below 10 keV would be completely hidden 
by the thick absorber, the observed emission being due to reflection 
from circumnuclear matter. In this scenario, the primary emission
could emerge above 10 keV (if the absorber is not too thick, see
e.g. Matt et al. 1999) and then be observable with the HXD/PIN.
As discussed in the previous section, the observed EWs of the iron
lines are consistent with the Compton-thick hypothesis (although in a 
quite unusual configuration), even if other
scenarios are also possible.  

\begin{figure}
\epsfig{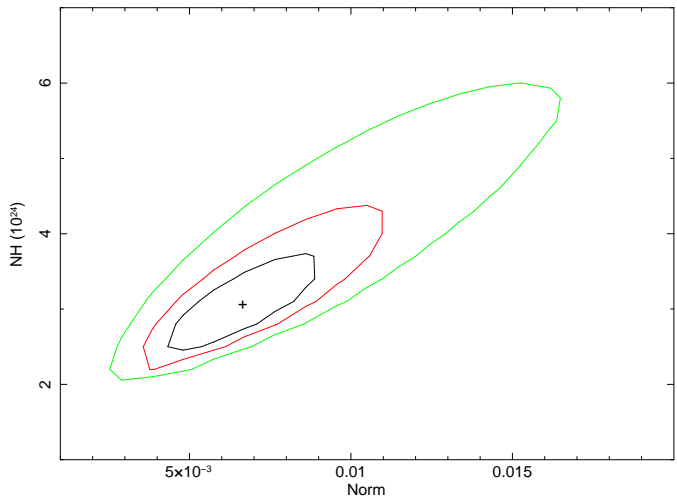}
  \caption{Contour plot of the putative Compton-thick absorber column density
vs. normalization of the primary power-law. 
}
  \label{contour_pin}
\end{figure}

In Fig.~\ref{pin} (left panel), the PIN spectrum is added to the already
analyzed XIS spectra, and
a clear excess with respect to the extrapolation of the XIS-only best fit
model is apparent (the result is basically the same if a high
temperature thermal plasma is used to account for the hydrogen-like 
iron line). No confusing source is known in the PIN field of view,
according to the existing catalogs of bright X-ray sources.
Assuming that the excess is due to the nuclear
radiation piercing through a Compton-thick absorber, we included 
in the model the transmission and scattering components expected in such a case
(Matt et al. 1999). In practice, we added to the model described in 
the previous section (which in this case should represent the nuclear
emission reflected mostly by highly ionized matter) the transmitted and scattered
(by the torus) emission using the MYTorus model (Murphy \& Yaqoob 2009; see also
http://www.mytorus.com/). 
The inclination angle of the system was
fixed to 90 degrees for simplicity. 
The values of the absorber column density
and of the normalization (at 1 keV) of the intrinsic radiation are strongly correlated,
and shown in Fig.~\ref{contour_pin}.  
The normalization of the unabsorbed (reflected in this scenario) power
law is about 4$\times$10$^{-4}$ in the same units, which implies that at the 90\% confidence
level the ratio between primary and reflected components ranges from about 10 to 30,
a value somewhat lower than usually found (e.g. Panessa et al. 2006, Marinucci et al. 2012).
The 15-100 keV flux of the source is 1.2$\times$10$^{-11}$ erg cm$^{-2}$ s$^{-1}$.
The 2-10/20-100 keV ratio is therefore more typical of Compton-thin sources, according
to the diagnostic diagram of Malizia et al. (2007). 

This result is only marginally consistent with the upper limit of 
1.3$\times$10$^{-11}$ erg cm$^{-2}$ s$^{-1}$ to the 20-100 keV flux obtained
by BeppoSAX (Dadina 2007) and with the upper limit of 7$\times$10$^{-12}$ erg cm$^{-2}$ s$^{-1}$
to the 20-40 keV flux obtained by INTEGRAL on September/October 2009 (this work).
It is inconsistent with the Swift-BAT observation of this source,
which provides only an upper limit to the 15-150 keV flux 
of  4$\times$10$^{-12}$ erg cm$^{-2}$ s$^{-1}$ (La Parola, private communication).
It must be recalled that the PIN spectrum is obtained adopting the standard model
for the background, which has an estimated average reproducibility of 
3\% at 1$\sigma$ (Fukazawa et al. 2009)\footnote{The cosmic X-ray background is also 
highly variable, up to 10\%, from place to place on scales of 1 sq degree (Barcons
et al. 2000). However, the CXB is only 5\% of the total PIN background (Fukazawa
et al. 2009).}. For individual observations, however, deviations of the background
as high as 5\% are sometimes observed (Pottschmidt, private communication). 
A background higher by 3.5\% suffices to
reduce the 15-100 keV flux to a value consistent with the Swift-BAT upper limit,
and in this case no excess is apparent (Fig.~\ref{pin}, right panel). Unfortunately,
this observation does not include a period of Earth occultation, so the real level
of the background cannot be determined.

\section{Conclusions}

Based on a short XMM-Newton observation, 
there were good arguments against the Compton-thick hypothesis
for NGC~3147 (Bianchi et al. 2008): the X-ray/[OIII] ratio is typical of 
unobscured objects, as well as the X-ray spectrum: in Compton-thick 
sources the 2-10 keV emission is usually much harder, being dominated by
the Compton reflection component from low ionization matter and with a neutral
iron line of about 1 keV EW.

However, if reflection is from highly ionized matter the spectrum is steeper, but
in this case emission from He- or H-like ions 
is also expected (Matt et al. 1996). Our $Suzaku$ 
observation suggests the presence of a strong H-like
line. Indeed, the overall line and continuum spectrum below 10 keV is consistent
with a Compton-thick scenario in which reflection is mostly (but not entirely)
due to a highly ionized mirror. Interestingly, if this is the case, the absorber may not be
the torus, which in this source is claimed by Shi et al. (2010) to be seen almost
face-on. Strong highly ionized reflectors in Compton-thick Seyfert 2s are 
rare, and even when observed, as in NGC~1068, their relative importance 
with respect to the low ionzation reflector is lower
(Iwasawa et al. 1997, Matt et. al. 2004). NGC~3147 would therefore be rather extreme.
It is worth noting that a larger-than-usual
amount of highly ionized reflection may at least partly explain the unusually (for a Compton-thick
source) high X-ray/[OIII] ratio, as well as the low primary-to-reflection ratio.

On the other hand, it must also be stressed that alternative hypotheses, like e.g.
large iron overabundance or a line emission from a very hot (temperature of about 20 keV),
optically thin plasma cannot be ruled out. These characteristics would be unusual 
in Seyfert galaxies, but NGC~3147, if confirmed as a ``true'' Seyfert 2 galaxy, 
would be a peculiar source anyway. 

A strong argument against the Compton-thick scenario 
is provided by the observed flux variation between the $Suzaku$ and the
XMM-$Newton$ observations, obtained 3.5 years apart. Flux variations 
on yearly time scales were also found comparing earlier observations  
with ASCA, BeppoSAX and Chandra. 
These variations indicate that we are looking at an emitting region with a size
smaller than a parsec. On the other hand, the argument, albeit strong,
is still not decisive. Variations
on similar or even smaller time scales have been claimed for NGC~1068, the archetypal 
Compton-thick Seyfert 2 (Guainazzi et al. 2000, Colbert et al. 2002). Interestingly, 
in NGC~1068 the variations seems to concern the highly ionized reflector, which in that source
is prominent only above about 2 keV (reflection from less ionized
matter, possibly related to the Narrow Line Regions, dominates at lower energies).
Moreover, there is evidence in some sources, most notably NGC~1365 (Risaliti et
al. 2005), of Compton-thick material very close to the central black hole.

If the source is moderately Compton-thick, we expect that the direct emission
would pierce through the absorber in hard X-rays. 
Using the standard background subtraction, an excess
in the PIN is observed, which however disappears - becoming consistent with 
the SWIFT/BAT upper limit - assuming that the background is higher
than usual by 3.5\%, a not uncommon variation. Therefore, this observation
is unfortunately not decisive in this respect. 

To conclude, after the new $Suzaku$ observation the 
``true'' Seyfert 2 nature of NGC~3147 
remains the most probable hypothesis. On the other hand, the ionized
iron line is best explained in the ``highly ionized reflector'' 
Compton-thick scenario, which is still a viable option. 
Future sensitive hard X-ray observations with {\it NuStar} and/or
high spectral resolution observations with {\it Astro-H} are needed
to definitely settle the issue.

\section*{Acknowledgements}
We thank the anonymous referee for her/his valuable suggestions.
Katja Pottschmidt is gratefully acknowledged for her help on 
the PIN background, and Valentina La Parola for providing the
SWIFT/BAT upper limit to the flux of the source.

GM, SB and FP acknowledge financial support from ASI under grants ASI/INAF
I/088/06/0 and I/009/10/0, FP also from grant INTEGRAL I/033/10/0.
XB's research is funded by the Spanish Ministry of Economy and Competitivity 
through grant AYA2010-21490-C02-01.

\end{document}